\documentclass[12pt,preprint]{aastex}

\begin{document}

\title{Variability of the High-Magnetic Field X-ray Pulsar PSR J1846$-$0258
Associated with the Supernova Remnant Kes 75 as Revealed by the Chandra
X-ray Observatory}
\author{Harsha Sanjeev Kumar \altaffilmark{1} 
\& Samar Safi-Harb\altaffilmark{1,2}}
\altaffiltext{1}{Department of Physics \& Astronomy, University of Manitoba, Winnipeg, MB R3T 2N2, Canada; harsha@physics.umanitoba.ca, samar@physics.umanitoba.ca}
\altaffiltext{2}{Canada Research Chair}

\begin{abstract}

We present results from the archival $Chandra$ observations of the 0.3~s X-ray pulsar PSR J1846$-$0258 associated with the supernova remnant (SNR) Kes 75. The pulsar has the highest spin-down luminosity ($\dot{E}$=8.3$\times$10$^{36}$ erg s$^{-1}$) among all the high magnetic field pulsars (HBPs) and has been classified as a Crab-like pulsar despite its magnetic field (5$\times$10$^{13}$ G) being above the quantum critical field. It is the only HBP described by a non-thermal Crab-like spectrum, powering a bright pulsar wind nebula (PWN). Our spectroscopic study shows evidence of spectral softening (photon index $\Gamma$=1.32$^{+0.08}_{-0.09}$ to 1.97$^{+0.05}_{-0.07}$) and temporal brightening (unabsorbed flux $F_{unabs}$=(4.3$\pm0.2)\times$10$^{-12}$ to 2.7$^{+0.1}_{-0.2}\times$10$^{-11}$ erg cm$^{-2}$ s$^{-1}$) of the pulsar by $\sim$6 times from 2000 to 2006. The 0.5--10 keV luminosity of the pulsar at the revised distance of 6 kpc has also increased from $L_{X}$=(1.85$\pm$0.08)$\times$10$^{34}$ to (1.16$^{+0.03}_{-0.07}$)$\times$10$^{35}$ erg s$^{-1}$, and the X-ray efficiency increased from 0.2$\pm$0.01$\%$ to 1.4$^{+0.04}_{-0.08}$$\%$. The observed X-ray brightening and softening of the pulsar suggests for the first time that this HBP is revealing itself as a magnetar.

\end{abstract}


\section{Introduction}

Magnetars are highly magnetized neutron stars with magnetic fields $B$$\sim$10$^{14-15}$ G, at least two orders of magnitude higher than the Crab-like pulsars ($B$$\sim$10$^{12}$ G). Two types of objects are thought to be magnetars -- the Anomalous X-ray Pulsars (AXPs) and the Soft Gamma Repeaters (SGRs), having spin periods $P$$\sim$2$-$12 s and X-ray luminosities, $L_X$, much larger than their spin-down luminosities, $\dot{E}$ (Woods \& Thompson 2006). There is now growing evidence of a new and small class of pulsars, the high magnetic field pulsars (HBPs), with spin and magnetic properties intermediate between the rotation-powered pulsars and the magnetars, and with magnetic fields $B\ge B_{QED}$; where $B_{QED}=m_e^2 c^3/\hbar e=4.4\times10^{13}$ G is the so-called quantum critical field. It is still under debate whether these HBPs form transient objects between the classical pulsars and the magnetars or whether they stand as a separate class of population. So far, there are six HBPs discovered with $B\ge B_{QED}$, 5 of which are discovered as radio pulsars (see Table 3 of Safi-Harb \& Kumar 2007 and references therein for a summary of their properties). 

PSR J1846$-$0258 is an energetic HBP discovered in X-rays, powering a bright pulsar wind nebula (PWN), at the center of the composite type SNR Kes 75 (SNR G29.7$-$0.3) of $\sim$ 3.5{\hbox{$^{\prime}$}} in diameter (Gotthelf et al. 2000). It has a rotation period $P$=324 ms, a period derivative $\dot{P}$=7.1$\times$10$^{-12}$ s s$^{-1}$, a magnetic field $B$=5$\times$10$^{13}$ G, and a spin-down luminosity $\dot{E}$=8.3$\times$10$^{36}$ erg s$^{-1}$. With a characteristic age of 723 yrs, it is likely the youngest of all known rotation-powered pulsars (Gotthelf et al. 2000).
No radio pulsations have been detected so far from this pulsar. It has been classified as a Crab-like pulsar because of its non-thermal hard X-ray spectrum and for powering a bright PWN (Gotthelf et al. 2000, Helfand et al. 2003 -- hereafter H03). Recent timing measurements using the \textit{Rossi X-ray Timing Explorer (RXTE)} placed an upper limit on the spin-down age of 884 yrs for the pulsar, the smallest estimated age for any rotation-powered pulsar based on a braking index $n$=2.65$\pm$0.01 (Livingstone et al. 2006). New \textit{Chandra} and \textit{Spitzer Space Telescope} observations of the SNR Kes 75 suggests a Wolf-Rayet progenitor for the supernova explosion (Morton et al. 2007). Until now, all calculations for the PSR J1846$-$0258 were based on a distance of 19 kpc obtained from neutral hydrogen absorption measurements (Becker \& Helfand 1984). Also, its  X-ray efficiency ($\eta_X$=$L_{X}$/$\dot{E}$=5$\%$; calculated using luminosities quoted by H03 at a distance of 19~kpc) is very high compared to rotation-powered pulsars. Recently, Leahy \& Tian (2008) obtained a more accurate distance of 6 kpc for the SNR Kes 75 using HI+$^{13}$CO observations of the VLA Galactic Plane Survey. Revision of the distance from 19 kpc to 6 kpc lowers the luminosity and the X-ray efficiency by a factor of 10.
$\gamma$-ray observations using $INTEGRAL$ detected soft $\gamma$-rays from the pulsar
and its PWN with a combined efficiency of $\sim$2.7$\%$ (recalculated using a distance of 6 kpc) in the $Chandra$ (2--10 keV) and $INTEGRAL$ (20--100 keV) bands (McBride et al. 2008).

PSR~J1846$-$0258's spin properties are remarkably similar to those of the radio pulsar PSR J1119$-$6127, however their X-ray properties are very different (see \S\ref{4.2}). Although PSR~J1846$-$0258 has been classified as a Crab-like pulsar, its spin properties and unusually high X-ray efficiency have long been suspected to be linked to its magnetar-strength field (e.g. Gotthelf et al. 2000).  In order to compare the properties of these two HBPs and shed light on the intriguing nature of PSR~J1846$-$0258, we analyzed four recent \textit{Chandra} archived observations. In this Letter, we report the results from this analysis and show the first evidence of variability from the pulsar, revealing an activity most likely linked to its magnetar-strength field.

\section{Observations \& Imaging Results}
\label{2}

PSR J1846$-$0258 was observed with the \textit{Advanced CCD Imaging Spectrometer (ACIS)} onboard the \textit{Chandra X-ray Observatory (CXO)} in 2000 Oct 15 (\textit{Obsid}: 748), and again in 2006 June 7-12 (\textit{Obsid}: 6686, 7337, 7338, 7339). The source was positioned at the aimpoint of the back-illuminated S3 chip of the ACIS. The CCD temperature was $-$120 C, with a CCD frame readout time of 3.2 s. The data were reduced using the standard CIAO 3.4 routines. The resulting effective exposure time was 192.2 ks for all the five observations. 

Figure 1 shows the combined ACIS-S3 image of the thermally emitting plasma from the SNR Kes 75 (studied in detail by Morton et al. 2007) and the PSR J1846$-$0258 ($\alpha_{J2000}$=18$^{h}$46$^{m}$24$^{s}$.94$\pm$0$^s$.01, $\delta_{J2000}$=$-$02$^{\circ}$58\hbox{$^{\prime}$}30\hbox{$^{\prime\prime}$}.1$\pm$0\hbox{$^{\prime\prime}$}.2; H03) surrounded by a bright and hard PWN. The background-subtracted images were divided in the soft (0.5-1.15 keV; red), medium (1.15-2.3 keV; green) and hard (2.3-10.0 keV; blue) energy bands, adaptively smoothed using a Gaussian with $\sigma$=1${\hbox{$^{\prime\prime}$}}$-3${\hbox{$^{\prime\prime}$}}$ for a significance of detection 2 to 5, and finally combined to produce the image shown in Fig. 1. A detailed spatially resolved spectroscopic study of the PWN features is beyond the scope of this paper. Here, we will focus on the overall spectrum of the PWN to study any changes in its properties from 2000 to 2006 in the context of the variability detected from the pulsar (see \S\ref{4.1}). 

\section{Spectroscopy of PSR J1846$-$0258 and its PWN}
\label{3}

The spectroscopic study of PSR J1846$-$0258 and its PWN was carried out in the 0.5--10 keV energy range using XSPEC version 12.4.1. The spectrum of the pulsar was extracted by selecting a circular region of 2{\hbox{$^{\prime\prime}$}}-radius centered on the pulsar (Fig. 1), encompassing 90$\%$ of the encircled energy\footnote{http://cxc.harvard.edu/proposer/POG/html/ACIS.html}. The background region was selected from an annular ring from 3{\hbox{$^{\prime\prime}$}} to 4{\hbox{$^{\prime\prime}$}} centered on the pulsar to subtract the contamination by the surrounding bright PWN. The spectrum of the PWN was extracted from a circular region of 18{\hbox{$^{\prime\prime}$}}-radius (Fig. 1). The background was extracted from a source-free region to the north-west of the pulsar. The spectra extracted were grouped by a minimum of 20 and 50 counts per bin for the 2000 and 2006 data, respectively, and the errors were calculated at the 90$\%$ confidence level.

A simple absorbed power-law (PL) model to the pulsar gave a poor fit due to CCD pileup\footnote{See http://space.mit.edu/$\%$7Edavis/papers/pileup2001.ps}
leading to the artificial flattening of the observed spectrum. PSR J1846$-$0258, being a bright source, is affected by pileup with an estimated pileup fraction of $\sim$10$\%$ for the 2000 data (H03) and 18$\%$ for the 2006 data. To account for pileup in our analysis, we included a convolution model (``pileup'') in XSPEC. The resulting model (Fig. 2(a), Table 1) produced an excellent fit. The spectra were further examined by adding a blackbody (BB) component to the model (after freezing the PL parameters) to account for any surface thermal emission from the pulsar. We obtained a BB temperature $kT$=0.74$^{+0.32}_{-0.05}$ keV ($T$=8.6$^{+3.7}_{-0.6}$$\times$10$^6$ K) and an unabsorbed thermal flux of (4.7$^{+7.5}_{-2.2})\times10^{-13}$ erg~cm$^{-2}$~s$^{-1}$ (0.5--10 keV), which is only $\sim$2$\%$ of the total non-thermal flux. However, the current statistics did not allow us to confirm or constrain the additional BB component. Using the BB component and assuming isotropic emission, we infer a  BB radius $R$=0.23$^{+0.18}_{-0.05}$ km, which is in agreement with the conventional polar cap radius R$_{pc}$$\sim$0.5$(\frac{P}{0.1s})^{-1/2}$=0.27 km. We also independently carried out the spectral analysis of the 2000 data by freezing $N_H$=3.96$\times$10$^{22}$ cm$^{-2}$ to compare with the values obtained by H03. Our observed photon index $\Gamma$=1.32$^{+0.08}_{-0.09}$ is in good agreement with that obtained by H03 ($\Gamma$=1.39$\pm$0.04). Due to the limited number of counts, we could not add a BB component to the 2000 data. The variability of the pulsar's spectrum from 2000 to 2006 is discussed in \S\ref{4.1}. The PWN is well fitted with an absorbed PL as summarized in Table~1. By comparing the 2000 and the 2006 spectral parameters, we note that the PWN's luminosity and photon index have increased by $\sim$11$^{+3}_{-4}$$\%$ and $\sim$3$^{+2}_{-1}$$\%$, respectively; however within error these changes are not significant.

\section{Discussion}
\subsection{X-ray variability of PSR J1846$-$0258}
\label{4.1}

The PL fit to the pulsar clearly shows a flux enhancement and a spectral softening from 2000 to 2006 (Fig. 2(a), Table 1). The corresponding increase observed in the luminosity by more than a factor of 6 provides evidence for brightening of the pulsar, suggesting for the first time an activity most likely associated with its ultra-high $B$-field. We now discuss the implications of the observed variations in the spectrum of PSR J1846$-$0258 considered as a Crab-like pulsar so far. Firstly, rotation-powered pulsars are known to be steady while PSR J1846$-$0258 has brightened by a factor of 6, a behavior seen in  magnetars. For example, the transient magnetar XTE J1810-197 brightened by 2 orders of magnitude in 2003 (Ibrahim et al. 2004); and the most recently discovered 2-s magnetar, 1E 1547.0$-$5408 (Camilo et al. 2007), had its X-ray flux vary by a factor of 7 between 1980 and 2006 (Gelfand \& Gaensler 2007, Halpern et al. 2007). Secondly, the pulsar's spectrum softened as it brightened ($\Gamma$ increased from 1.32 to 1.97), indicating a spectral change that could be tied to a change in its $B$-field configuration or a magnetar-like burst. Thirdly, there is a hint that the PWN's spectrum has also changed as seen from the increase in its $L_X$ by $\sim$11$\%$ and a slight softening of its spectrum. While these changes are not significant within error, they support the emergence of a magnetar-like burst from PSR~J1846$-$0258 which would have injected relativistic particles in the PWN, thus enhancing its brightness observed in 2006. Furthermore, as pointed out by Kargaltsev \& Pavlov (2008), it is possible that the unusually high efficiency of the PWN could be explained by a series of magnetar-like bursts. Therefore, based on the observed changes, we conclude that PSR~J1846$-$0258 revealed its identity as a magnetar, likely after a magnetar-like burst.
 
Like the young rotation-powered pulsars, PSR J1846$-$0258's X-ray spectrum is described by a PL spectrum with a hard photon index, even at its steeper phase ($\Gamma$=1.97). The only 2 magnetars having hard PL X-ray spectra with $\Gamma$$\sim$2 are SGR~1806$-$20 and SGR~1900+14 (Woods \& Thompson, 2006). As well, its X-ray efficiency, despite being high in comparison to rotation-powered pulsars, is still smaller than 1 -- a characteristic that differentiates rotation-powered pulsars from magnetars. Although these properties pose some difficulty in interpreting the pulsar as a newly emerging magnetar, they can be attributed to the pulsar being young and highly energetic. In particular, magnetars' X-ray luminosities can easily exceed their spin-down luminosity because they have a much lower $\dot{E}$ ($\sim$10$^{33}$ erg s$^{-1}$), ranging from $\sim$6$\times$10$^{31}$~erg~s$^{-1}$ for 1E~2259+586 to 5$\times$10$^{34}$~erg~s$^{-1}$ for SGR~1806$-$20. The highest known $\dot{E}$ of a magnetar is that of the 2-s magnetar, 1E 1547.0$-$5408, with $\dot{E}$=1.0$\times$10$^{35}$~erg~s$^{-1}$ (Camilo et al. 2007), still $\sim$2 orders of magnitude smaller than that of PSR J1846$-$0258 . The X-ray luminosity of 1E 1547.0$-$5408 only just exceeded its $\dot{E}$ during outburst. Furthermore, magnetars are known to spin down fast to a period of a few seconds in $\leq$10,000 yrs, but PSR J1846$-$0258 is only $\leq$900 years and hence, its $\dot{E}$ ($\propto$$P^{-3}$) is still high. Therefore, based on pure energetic grounds, PSR J1846$-$0258's rotational energy loss can still power its $L_X$ and the surrounding PWN, like the rotation-powered pulsars. On the other hand, the unusually high X-ray efficiency (of the pulsar and PWN) and the variability observed between 2000 and 2006 suggests that at least a fraction of the X-ray luminosity is powered by magnetic energy that would be supplied e.g. during a magnetar-like burst. Finally we note that we cannot rule out the possibility that the pulsar's $L_X$ exceeded its $\dot{E}$ prior to the 2006 $Chandra$ observations during a magnetar-like burst that was missed by the $Chandra$ observations.

\subsection{Comparison with the HBP PSR~J1119$-$6127}
\label{4.2}

Of all the HBPs, PSR J1846$-$0258 has the highest $\dot{E}$=8.3$\times$10$^{36}$ erg s$^{-1}$ (Gotthelf et al. 2000) and $\eta_{X}$=1.4$\%$ (using the 2006 flux from this work). It has spin parameters similar to the HBP PSR~J1119$-$6127 ($P$=407 ms, $\dot{P}$=4$\times$10$^{-12}$ s s$^{-1}$, B=5$\times$10$^{13}$ G; Camilo et al. 2000) and both pulsars are now believed to be at about the same distance, $D$$\sim$6 kpc (Gonzalez \& Safi-Harb 2003, Leahy \& Tian 2008). Despite the remarkable similarities between their spin properties, they have very different X-ray properties. The resolved spectrum of PSR~J1119$-$6127 is best fitted by a two-component BB+PL model, with the BB component dominating the X-ray emission (Safi-Harb \& Kumar 2007), and characterized by a high pulsed fraction of 74$\pm$14$\%$ in the 0.5--2.0 keV (Gonzalez et al. 2005); however, the emission from PSR~J1846$-$0258 is dominated by a PL model with a hard photon index. Furthermore, the X-ray efficiency of PSR J1119$-$6127 is only $\sim$0.05$\%$ (Safi-Harb \& Kumar 2007), much lower than that of PSR J1846$-$0258. Moreover, PSR J1119$-$6127 powers a weak PWN ($\eta_{X}\sim$0.02$\%$, Safi-Harb \& Kumar 2007) while PSR J1846$-$0258 powers a much brighter PWN ($\eta_{X}\sim$1.85$\%$). While the data presented here show the first evidence of spectral changes from PSR~J1846$-$0258 between 2000 and 2006, no changes were observed in the $Chandra$ data of PSR J1119$-$6127 taken in 2002 \& 2004 (Safi-Harb \& Kumar 2007). These differences indicate that the B-field strength is not the sole parameter that determines their characteristics. We attribute these differences mainly to their evolutionary state (PSR~J1119$-$6127 being twice as old and less energetic), their different $N_H$ (PSR J846-0258's $N_H$ is at least twice as high as PSR J1119$-$6127's thus burying any soft BB component), and likely different progenitor stars. Morton et al. (2007) suggests that Kes 75 could have resulted from the rare explosion of a Wolf-Rayet star during a Type Ib or Ic event. Furthermore, PSR~J1846$-$0258 has revealed its magnetar-like nature only now. Since two transient magnetars have now been observed in the radio (Camilo et al. 2006, 2007), monitoring of PSR~J1846$-$0258 in the radio and other wavelengths will shed more light on its magnetar nature. As well, monitoring of PSR~J1119$-$6127 in the X-ray may reveal in the future a magnetar-like activity as observed for PSR J1846$-$0258, and as recently inferred for 1E~1547.0-5408 (Gelfand \& Gaensler 2007) .
  
\section{Conclusions}

We have used the archival $Chandra$ data to carry out a spectroscopic study of the HBP PSR~J1846$-$0258, classified as a Crab-like pulsar, and its PWN in the 0.5--10 keV energy band. Our analysis of the pulsar's spectrum revealed flux enhancement by a factor of more than 6 and X-ray softening from 2000 to 2006, accompanied by a hint of flux enhancement of the PWN. This provides the first observational evidence of a magnetar-like emission associated with this HBP. It is still intriguing that the X-ray luminosity, even at its observed peak, did not exceed the pulsar's rotational energy power, suggesting that spin-down can energetically power the X-ray emission from the pulsar and its PWN. Further observations of this source at all wavelengths and monitoring of other HBPs will answer the question whether HBPs are indeed transient magnetars disguised in other forms, and will shed more light on the nature of these sources and their link to magnetars and the rotation-powered pulsars.

\acknowledgements
We acknowledge a recent private communication with M. Gonzalez who, prior to the submission of this paper, made us aware of the Gavriil et al. (2008) finding (paper posted on the arXiv after submission of our manuscript), which confirm the results and conclusions we had reached independently prior to the conversation, and which further establish the magnetar nature of the pulsar. It was also brought to our attention, after submission of our paper, that the variability detected from the pulsar with $Chandra$ was also independently found by Ng et al. 2008 
(P. Slane, private communication). We thank F. Camilo for commenting on our manuscript. S. Safi-Harb acknowledges support by the Natural Sciences and Engineering Research Council of Canada and the Canada Research Chairs program. This research made use of NASA's Astrophysics Data System and of NASA's HEASARC maintained at the Goddard Space Flight Center.

\clearpage

\begin{table*}[ht]
\caption{Spectral fits to the PSR J1846$-$0258 and its PWN for the 2000 and 2006 observations}
\small
\begin{tabular}{l l l l l l}
\hline\hline
Parameter & \multicolumn{2}{c}{2006 PSR} & 2000 PSR & 2006 PWN & 2000 PWN\\
\cline{2-3}
& PL & BB+PL & PL & PL & PL\\
\hline

$ N_{H} $(10$^{22}$ cm$^{-2}$)$^a$ & 4.15$_{-0.12}^{+0.09}$ & 4.15 (frozen) & 3.96 (frozen) & 3.92$\pm$0.05 & 3.84$_{-0.12}^{+0.10}$\\

$\Gamma$ & 1.97$_{-0.07}^{+0.05}$& 1.97 (frozen) & 1.32$_{-0.09}^{+0.08}$ & 1.89$\pm$0.03 & 1.83$_{-0.06}^{+0.05}$\\

$kT$ (keV) & \nodata & 0.74$_{-0.05}^{+0.32}$ & \nodata & \nodata &\nodata \\

$F$ (PL)$^b$  & 2.7$_{-0.2}^{+0.1}\times10^{-11}$ & 2.4$_{-0.3}^{+1.3}\times10^{-11}$ & (4.3$\pm$0.2)$\times10^{-12}$ & 3.6$_{-0.1}^{+0.2}\times10^{-11}$ & 3.3$_{-0.1}^{+0.2}\times10^{-11}$\\

$F$ (BB)$^b$  &\nodata & 4.7$_{-2.2}^{+7.5}\times10^{-13}$ & \nodata &\nodata &\nodata\\

$\chi_{\nu}^2~(dof)$ & 1.02 (833) & 1.01 (832) & 1.10 (165)& 0.99 (1170) & 0.91 (387) \\

$L_{X}$$^c$ & 1.16$^{+0.03}_{-0.07}\times$10$^{35}$ & 1.03$^{+3.40}_{-0.96}\times$10$^{35}$& (1.85$\pm$0.08)$\times$10$^{34}$ & 1.55$_{-0.07}^{+0.05}\times10^{35}$ & 1.4$_{-0.05}^{+0.09}\times10^{35}$\\

$\eta_{X}$ &1.4$^{+0.04}_{-0.08}$$\%$ & 1.2$^{+4.0}_{-1.1}$$\%$ & 0.2$\pm$0.01$\%$ &1.85$^{+0.06}_{-0.08}\%$ & 1.6$^{+0.10}_{-0.05}\%$ \\ \hline
\end{tabular}
\tablecomments{$^a$The derived column density from the XSPEC `wabs' model which uses the Wisconsin cross-sections (Morrison \& McCammon 1983). $^b$ Unabsorbed flux (0.5--10 keV) in units of erg cm$^{-2}$ s$^{-1}$. $^c$ Luminosity (0.5--10 keV) in units of erg s$^{-1}$.}
\end{table*}

\clearpage

\begin{figure}[h]
\center
\plotone{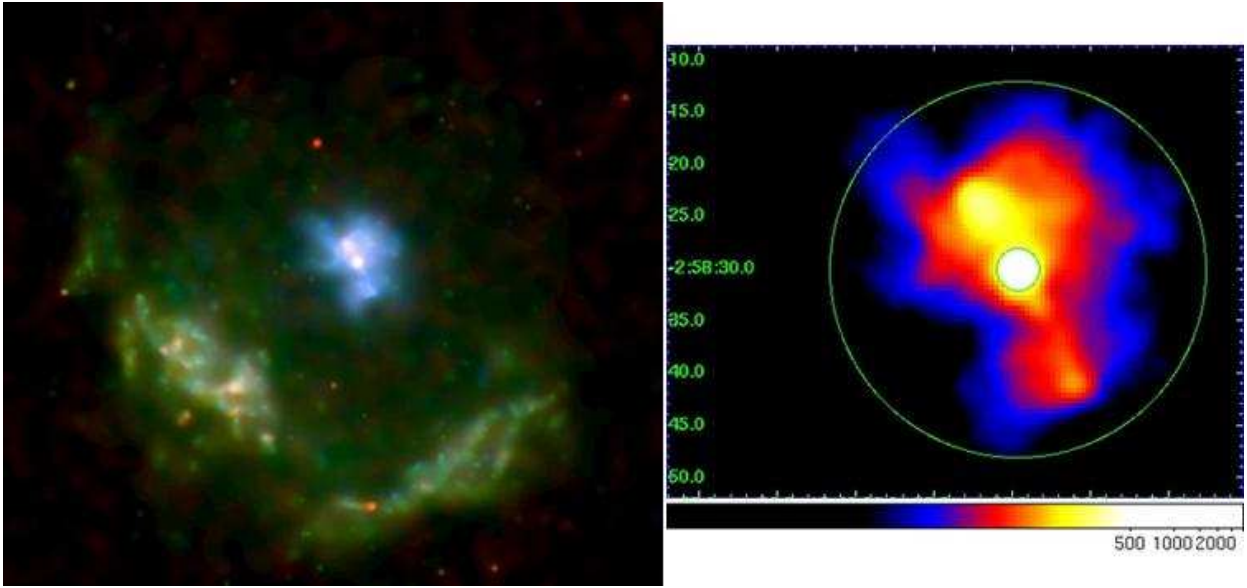}
\caption{$Left$: $Chandra$ ACIS-S3 tri-color image of the PSR J1846$-$0258 and its hard PWN at the center of the SNR Kes 75 (see \S\ref{2}). $Right$: The intensity image of PSR J1846$-$0258 and its PWN (in logarithmic scale) adaptively smoothed using a Gaussian with $\sigma$=1\hbox{$^{\prime\prime}$}-3\hbox{$^{\prime\prime}$} for a significance of detection 2 to 5. Overlaid in green are the regions selected for the spectroscopic study of the pulsar (2\hbox{$^{\prime\prime}$}-radius central circle) and PWN (18\hbox{$^{\prime\prime}$}-radius outer circle excluding the pulsar) (see \S\ref{3}).}
\end{figure}

\begin{figure*}[ht]
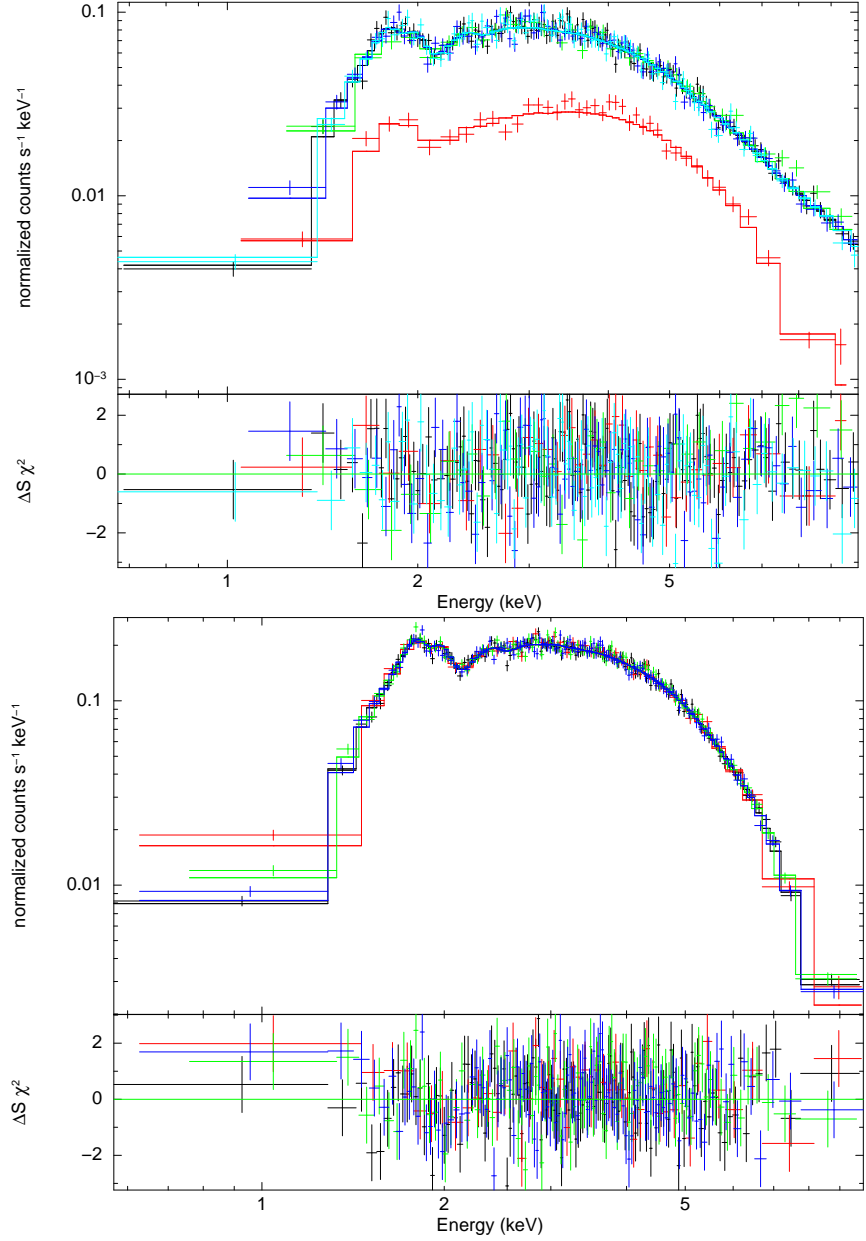
 
\center
\includegraphics[width=0.5\textwidth, angle=-90]{f2a.eps}{\label{figure:psr}}
\includegraphics[width=0.5\textwidth, angle=-90]{f2b.eps}{\label{figure:pwn}}
\caption{\textit{(a)} The 2000 (red crosses) and 2006 data of PSR~J1846$-$0258 fitted with an absorbed PL model, corrected for pileup.
\textit{(b)} The 2006 PWN data fitted with an absorbed PL model. 
\textit{Bottom panels}: Residuals of the fit to an absorbed PL model.}
\end{figure*}

\end{document}